\documentstyle[emulateapj,flushrt]{article}
\newcommand{\beq}{\begin{equation}}
\newcommand{\eeq}{\end{equation}}
\newcommand{\erg}{\textrm{ erg}}
\newcommand{\cm}{\textrm{ cm}}
\newcommand{\gauss}{\textrm{ gauss}}
\newcommand{\s}{\textrm{ s}}
\newcommand{\muas}{\ \mu {\rm as}}
\begin{document}
\title{Generation of Magnetic Fields in the Relativistic Shock of
Gamma-Ray-Burst Sources}

\author{Mikhail V. Medvedev\altaffilmark{1} and Abraham Loeb\altaffilmark{2}} 

\affil{Harvard-Smithsonian Center for Astrophysics, 60 Garden Street,
Cambridge, MA 02138}

\altaffiltext{1}{Also at the Institute for Nuclear Fusion, RRC ``Kurchatov
Institute'', Moscow 123182, Russia; E-mail: mmedvedev@cfa.harvard.edu; URL:
http://cfa-www.harvard.edu/\~{ }mmedvede/ }

\altaffiltext{2}{E-mail: aloeb@cfa.harvard.edu; URL:
http://cfa-www.harvard.edu/\~{ }loeb/ }

\begin{abstract}

We show that the relativistic two-stream instability can naturally generate
strong magnetic fields with $10^{-5}$--$10^{-1}$ of the equipartition
energy density, in the collisionless shocks of Gamma-Ray-Burst (GRB)
sources. The generated fields are parallel to the shock front and
fluctuate on the very short scale of the plasma skin depth. The synchrotron
radiation emitted from the limb-brightened source image is linearly polarized 
in the radial direction relative to the source center. Although the net
polarization vanishes under circular symmetry, GRB sources should exhibit
polarization scintillations as their radio afterglow radiation gets
scattered by the Galactic interstellar medium. Detection of polarization
scintillations could therefore test the above mechanism for magnetic field
generation.

\end{abstract}

\section{Introduction}

Cosmological $\gamma$-ray bursts (GRB) are believed to be produced in the
fireballs of very energetic explosions, when a large amount of energy,
$E\sim10^{51-54}\erg$, is released over a few seconds in a small volume 
in with a negligible baryonic load, $Mc^2\ll E$
(see Piran 1999 for a review). Most of the energy is eventually transferred
to the baryons which are accelerated to ultra-relativistic velocities with
a Lorentz factor $\gamma\simeq E/Mc^2\sim10^2$--$10^3$ (e.g., \cite{ShP90};
\cite{Paczynski90}). A substantial fraction of the kinetic energy of the
baryons is transferred to a non-thermal population of relativistic
electrons through Fermi acceleration at the shock (\cite{MR93}). The
accelerated electrons cool via inverse Compton scattering and synchrotron
emission in the post-shock magnetic fields and produce the radiation
observed in GRBs and their afterglows (e.g., \cite{Katz94}; \cite{SNP96};
\cite{Vietri97}; \cite{Waxman97}; \cite{WRM97}). The shock could be either
{\it internal} due to collisions between fireball shells caused by source
variability (Paczy\'nski \& Xu 1994; \cite{RM94}), or {\it external} due to
the interaction of the fireball with the surrounding interstellar medium
(ISM; \cite{MR93}). The radiation from internal shocks can explain the
spectra (Pilla \& Loeb 1998) and the fast irregular variability of GRBs
(Sari \& Piran 1997a), while the synchrotron emission from the external
shocks provides a successful model for the broken power-law spectra and
smooth temporal behavior of afterglows (e.g., Waxman 1997a,b).  In both
cases, strong magnetic fields are required behind the shocks at all times
in order to fit the observational data.

The properties of the synchrotron emission from GRB shocks are determined
by the magnetic field strength, $B$, and the electron energy distribution
behind the shock. Both of these quantities are difficult to estimate from
first principles, and so the following dimensionless parameters are often
used to incorporate modeling uncertainties (\cite{SNP96}), \beq
\epsilon_B\equiv\frac{U_B} {e_{\rm th}} ,\qquad
\epsilon_e\equiv\frac{U_e}{e_{\rm th}} . \eeq Here $U_B={B^2}/{8\pi}$ and
$U_e$ are the magnetic and electron energy densities and $e_{\rm
th}=nm_pc^2(\bar\gamma_p-1)$ is the total thermal energy density behind the
shock; where $m_p$ is the proton mass, $n$ is the proton number density,
and $\bar\gamma_p$ is the mean thermal Lorentz factor of the protons.  The
observed afterglow spectra and lightcurves typically yield values of the
magnetic energy parameter ranging from $\epsilon_B\sim0.1$ (Waxman 1997;
Wijers \& Galama 1998), down to $10^{-2}$ (Granot et al. 1998) or even
$\epsilon_B\sim10^{-5}$ (Galama et al. 1999; \cite{Vreeswijk99}) --- all
below the {\em equipartition} limit $\epsilon_B\sim 1$.

The existence of strong magnetic fields is naturally expected in the
compact environments of potential GRB progenitors. First, the field might
originate from a highly magnetized stellar remnant, such as a neutron star,
with $B\la 10^{16}\gauss$. Second, a turbulent magnetic dynamo could
amplify a relatively weak seed magnetic field in the vicinity of the
progenitor.  This process, however, requires the turbulence to be
anisotropic and have a nonzero total helicity, ${\bf v\cdot(\nabla\times
v)}\not=0$. A similar mechanism, called the $\alpha$-$\Omega$-dynamo, might
operate in rapidly rotating objects (\cite{Thompson94}, see also
\cite{MLR93}). Finally, the magnetic shearing instability
(\cite{BalbusHawley91}) could amplify the magnetic field (but not the flux)
in strongly sheared flows. The $e$-folding time for this instability is
approximately the rotation period, which decreases with radius as $R^{-2}$
due to angular momentum conservation in the outflowing wind. Thus, being
possibly important in the early stages of the fireball expansion
(\cite{NPP92}), this instability is inefficient at large radii, where its
$e$-folding time greatly exceeds the dynamical time of the fireball.

In contrast to such progenitor environments where large magnetic fields are
natural, there is currently no satisfactory explanation for the origin of
the strong magnetic fields required in GRB shocks (see discussions in
\cite{Thompson94}; \cite{MLR93}; \cite{SNP96}).  Compression of the ISM
magnetic field in external shocks yields a field amplitude $B\sim\gamma
B_{\rm ISM}\sim 10^{-4} (\gamma/10^2) \gauss$, which is too weak
(\cite{SNP96}) compared to the required equipartition value $B_{eq}\sim 50
(\gamma/10^2)(n_{\rm ISM}/1~{\rm cm^{-3}})^{1/2} \gauss$, and can account
only for $\epsilon_B=(B/B_{eq})^2\la 10^{-11}$.  Alternatively, some
magnetic flux might originate at the GRB progenitor and be carried by the
outflowing fireball plasma (or by a precursor wind). Because of flux
freezing, the field amplitude would decrease as the wind expands. In this
case, only a progenitor with a rather strong magnetic field $\sim
10^{16}\gauss$ might produce sufficiently strong fields during the GRB
emission. However, since the field amplitude scales as $B\propto V^{-2/3}$
for an expanding shell of volume $V$, even a highly magnetized plasma at
$R\sim 10^7\cm$ would possess only a negligible field amplitude of $\sim
10^{-2}\gauss$, or $\epsilon_{B}\la 10^{-7}$, at a radius of $R\ga
10^{16}\cm$, where the afterglow radiation is emitted\footnote{Both the
magnetic field energy density and the thermal energy of the fireball scale
as $\propto V^{-4/3}$ for adiabatic expansion.  However, when shocks are
generated, the plasma is heated due to the dissipation of the fireball
kinetic energy, and the magnetic energy parameter decreases far below
equipartition in the post shock gas.}  (see also \cite{MLR93}).  Moreover,
the emitting material behind the external shock is continuously replenished
by the ISM, and so the field originally carried by the fireball ejecta
cannot account for the afterglow radiation.

None of the above mechanisms is capable of generating near equipartition
magnetic fields in the {\em external shocks} which produce the delayed
afterglow emission.  In this paper, we propose a different, {\em universal}
mechanism of magnetic field generation in GRB shocks.  It involves the
relativistic generalization of the two-stream (Weibel) instability in a
plasma. This instability is driven by the {\em anisotropy} of the Particle
Distribution Function (PDF) and, hence, could operate in both internal and
external shocks.  Our main results are as follows:
\begin{enumerate}
\item
The characteristic $e$-folding time in the shock frame for the instability
is $\sim10^{-7}\s$ for internal shocks and $10^{-4}\s$ for external shocks.
This time is much shorter than the dynamical time of GRB fireballs.
\item
The generated magnetic field is randomly oriented in space, but always lies
in the plane of the shock front.
\item
The instability is powerful. It saturates only by nonlinear effects when
the magnetic field amplitude approaches equipartition with the electrons
(and possibly with the ions).
\item
The instability isotropizes the PDF and, thus, effectively heats the electrons 
and protons.
\item
The characteristic coherence scale of the generated magnetic field is of
the order of the relativistic skin depth, i.e. $\sim10^3\cm$ for internal
shocks and $\sim10^5\cm$ for external shocks. This scale is much smaller
than the spatial scale of the source.
\item
The mean free path for Coulomb collisions is larger than the fireball
size. However, the randomness of the generated magnetic field provides
effective collisions due to pitch angle scattering of the particles in an
otherwise collisionless plasma and, thus, justifies the use of the
magneto-hydrodynamic (MHD) approximation for GRB shocks. The magnetic
fields communicate the momentum and pressure of the outflowing fireball
plasma to the ambient medium and define the shock boundary.
\end{enumerate}

The above mechanism results in tangential magnetic fields near the apparent
limb of the source. Hence, the long-term synchrotron emission from the limb
would be linearly polarized along the radial direction relative to the
source center.  Although the net polarization of a circularly symmetric
source is zero, scattering of the radio afterglow emission of GRBs by the
intervening Galactic interstellar medium would break the symmetry in the
source image and result in polarization scintillations.  This effect can be
used to test the reality of our proposed mechanism for the generation of
magnetic fields in GRB blast waves.

The outline of the paper is as follows. The physical mechanism of the
instability is discussed in \S \ref{S:WEIBEL}.  The generation of magnetic
fields in internal and external shocks is discussed in \S
\ref{S:SHOCKS}. In \S \ref{S:PREDICT} we predict the polarization
scintillation signal in our model.  Finally, \S \ref{S:DISC} summarizes our
main conclusions.

\section{The Two-Stream Instability \label{S:WEIBEL}}

The instability under consideration was first predicted by Weibel (1959) for a
non-relativistic plasma with an anisotropic distribution function. The simple
physical interpretation provided later by Fried (1959) treated the PDF
anisotropy more generally as a two-stream configuration of a cold plasma. 
Below we give a brief, qualitative description of this two-stream magnetic 
instability.

\begin{figure}
\plotone{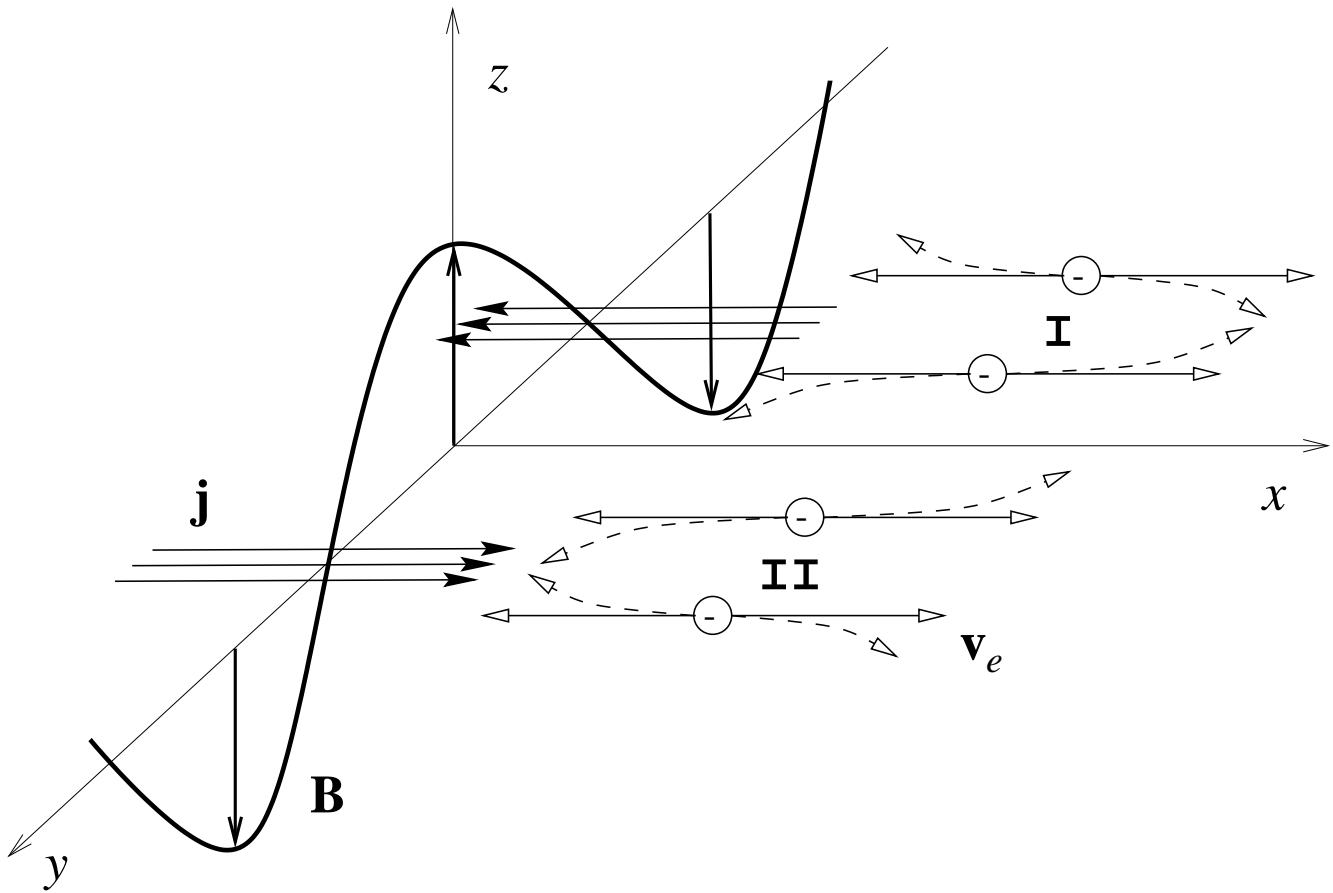}
\vskip1cm
\figcaption[f1.eps]{Illustration of the instability. A magnetic field
perturbation deflects electron motion along the $x$-axis, and results in
current sheets ($j$) of opposite signs in regions I and II, which in turn
amplify the perturbation. The amplified field lies in the plane
perpendicular to the original electron motion.
\label{fig} } 
\end{figure}

Let us consider, for simplicity, the dynamics of the electrons only, and
assume that the protons are at rest and provide global charge neutrality.
The electrons are assumed to move along the $x$-axis (as illustrated in
Figure\ \ref{fig}) with a velocity ${\bf v}=\pm {\bf \hat x}v_x$ and equal
particle fluxes in opposite directions along the $x$-axis (so that the net
current is zero). Next, we add an infinitesimal magnetic field fluctuation,
${\bf B}= {\bf\hat z}B_z\cos(ky)$. The Lorentz force, $-e{{{\bf v}\over
c}\times{\bf B}}$, deflects the electron trajectories as shown by the
dashed lines in Figure\ \ref{fig}. As a result, the electrons moving to the
right will concentrate in layer I, and those moving to the left -- in
layer II. Thus, current sheaths form which appear to {\em increase} the
initial magnetic field fluctuation. The growth rate is $\Gamma=\omega_{\rm
p}v_y/c$, where $\omega_{\rm p}^2=(4\pi e^2n/m)$ is the non-relativistic
plasma frequency (\cite{Fried59}). Similar considerations imply that
perpendicular electron motions along $y$-axis, result in oppositely
directed currents which suppress the instability. The particle motions
along ${\bf\hat z}$ are insignificant as they are unaffected by the
magnetic field. Thus, the instability is indeed driven by the PDF
anisotropy and should quench for the isotropic case.

The Lorentz force deflection of particle orbits increases as the magnetic
field perturbation grows in amplitude. The amplified magnetic field is
{\em random} in the plane perpendicular to the particle motion, since it is
generated from a random seed field. Thus, the Lorentz deflections result in
a pitch angle scattering which makes the PDF isotropic. If one starts from a
strong anisotropy, so that the thermal spread is much smaller than the
particle bulk velocity, the particles will eventually isotropize and the
thermal energy associated with their random motions will be equal to their
initial directed kinetic energy. This final state will bring the
instability to saturation.

We note the following points about the nature of the instability:
\begin{enumerate} 
\item
The instability is {\em aperiodic}, i.e., ${\rm Re}\,\omega=0$. Thus, it
can be saturated by nonlinear effects only, and not by kinetic effects such
as collisionless damping or resonance broadening. Hence, the magnetic field
can be amplified up to high values.

\item
Despite its intrinsically kinetic nature, the instability is 
non-resonant,\footnote{This instability may be treated as an analog 
of the fire-hose instability in the absence of the external magnetic field.}
i.e., it is impossible to single out a group of particles that is
responsible for the instability. Since the bulk of the plasma participates
in the process, the energy transferred to the magnetic field could be
comparable to the total kinetic energy of the plasma. Hence, the
instability is {\em powerful}.

\item
The instability is self-saturating. It continues until all the free energy
due to the PDF anisotropy is transferred to the magnetic field energy.

\item
The generated magnetic field always lies in the plane perpendicular to the
initial anisotropy axis of the PDF, i.e., to the shock propagation
direction.

\item
The produced magnetic field is randomly oriented in the shock plane.  The
Lorentz forces randomizes particle motion over the pitch angle and, hence,
introduces an effective scattering process into the otherwise collisionless
system.  This validates the use of the MHD approximation in the study of
collisionless GRB shocks.

\end{enumerate}

\cite{SGbook} and \cite{MS63} provide a kinetic, non-relativistic treatment
of the instability in both the linear and the quasi-linear regimes, and
apply the theory of collisionless shocks to space plasmas.  In the next
section we will extend their analysis to the case of ultra-relativistic GRB
shocks.

\section{Magnetic Field Generation in GRB Shocks \label{S:SHOCKS}}

We consider a GRB shock front expanding at a Lorentz factor, $\gamma_{\rm
sh}$, behind which the particles have a thermal Lorentz factor,
$\bar\gamma$.  In this section, we will derive equations which are equally
applicable to electrons and protons, whichever species dominates the growth
of the instability.  Later, we shall use the subscripts ``$e$'' and ``$p$''
to denote electrons and protons, respectively. We calculate all quantities
in the comoving frame of the shock.

A fully kinetic, relativistic treatment of the magnetic two-stream (Weibel)
instability is a complicated task. The dispersion relation for a simplified
``water-bag'' PDF was derived by Yoon \& Davidson (1987) and is given by
equation\ (\ref{disp}) in Appendix \ref{A1}. This dispersion relation
implies that only a range of modes above a critical wavelength will grow
[cf. Eq.\ (\ref{range})]. Naturally, the mode with the largest growth rate,
$\Gamma_{\rm max}$, dominates and sets the characteristic length-scale of
the magnetic field fluctuations, $\lambda\sim k_{\rm max}^{-1}$. The
ultra-relativistic expressions for $\Gamma_{\rm max}$ and $k_{\rm max}$ are
given by equation\ (\ref{gamma-k}) for a strong initial anisotropy. We
write the corresponding $e$-folding time and correlation length of the
field as \beq \tau\simeq\frac{\gamma_{\rm sh}^{1/2}}{\omega_{\rm p}} ,\quad
\lambda\simeq2^{1/4}\frac{c\bar\gamma^{1/2}}{\omega_{\rm p}} .
\label{scales} 
\eeq

We can now estimate the nonlinear saturation amplitude of the magnetic
field. The instability is due to the free streaming of particles. As the
field amplitude grows, the transverse deflection of particles gets
stronger, and their free streaming across the field lines is
suppressed. The typical curvature scale for the deflections is the Larmor
radius, $\rho=v_{\bot B}/\Omega_{\rm c}\simeq(\gamma_{\bot
B}^2-1)^{1/2}mc^2/eB$, where $v_{\bot B}$ and $\gamma_{\bot B}$ are the
transverse velocity and Lorentz factor of a particle relative to the local
magnetic field.  On scales larger than $\rho$, particles can only move
along field lines. Hence, when the growing magnetic fields become such that
$k_{\rm max}\rho\sim1$, the particles are magnetically trapped and can no
longer amplify the field. Assuming an isotropic particle distribution at
saturation ($\gamma_{\bot B}\sim\bar\gamma$), this condition can be
re-written as \beq \frac{B^2/8\pi}{mc^2n(\bar\gamma-1)}
\sim\frac{(\bar\gamma+1)}{2\sqrt2\,\bar\gamma} .
\label{sat}
\eeq For ${\bar\gamma}\gg 1$, this corresponds to a magnetic energy density
close to equipartion with the amplifying particles.  Interestingly, one may
obtain the same result following a different analysis.  First, the
instability leads to a growth of the field amplitude [as given by the last
term in Eq.\ (\ref{ke}), $\sim{\bf v\cdot}\partial_{\bf x} f$]. Second,
nonlinearity leads to the transfer of energy to shorter wave lengths,
$k>k_{\rm crit}$, where the fluctuations are damped [as described by the
second term in Eq.\ (\ref{ke}), $\sim(e/c) {\bf v\times
B\cdot}\partial_{\bf p} f$]. Thus, the steady value of $B$ is determined by
balancing these two processes.  Equating these two terms and replacing
$\partial_x$ by $k_{\rm crit}\simeq k_{\rm max}/\sqrt2
\sim\rho_p/\sqrt{2}$, yields \beq v_{\bot B} k_{\rm crit}f\sim v_{\bot
B}eBf/mc^2\gamma_{\bot B} .  \eeq The field strength estimated here is
equivalent to that given in equation\ (\ref{sat}) to within a factor of
order unity.

Direct computer simulations of the instability in both non-relativistic and
relativistic electron plasmas confirm that the saturation occurs at
slightly {\em sub-equipartition} values of $B$ (see, e.g.,
\cite{Califanoetal98}; Kazimura et al. 1998; \cite{Yangetal94};
\cite{WE91}), 
\beq \frac{B^2/8\pi}{mc^2n(\bar\gamma-1)}\equiv\eta\sim0.01 - 0.1\, .
\label{field}
\eeq 
where we introduced the efficiency factor $\eta\lesssim0.1$.  The precise
saturation level depends on the nonlinear modification of the PDF during
the instability which is not accounted for by our linear analysis.  We
shall retain the efficiency factor, $\eta$, in our estimates.

Note that the thermal Lorentz factor of particles, $\bar\gamma$, varies in
time as the instability develops. Due to particle scattering by the
generated magnetic fields, an initially highly anisotropic PDF with
$\bar\gamma\ll\gamma_{\rm sh}$ will eventually evolve to an isotropic,
ring-like distribution, for which $\bar\gamma\simeq\gamma_{\rm sh}$. Thus,
the spatial scale and amplitude of the resultant magnetic field, given by
equations\ (\ref{scales}) and (\ref{field}), will evolve during the
lifetime of the instability because they are functions of $\bar\gamma$. In
estimating these values at a GRB shock when the instability saturates, we
take $\bar\gamma\simeq\gamma_{\rm sh}$.  The $e$-folding time for the
instability is independent of $\bar\gamma$ in the case of strong
anisotropy. In the case of weak anisotropy, $\bar\gamma\approx\gamma_{\rm
sh}$, the ``water-bag'' model used here is formally invalid, but the
comparison of the ultra-relativistic results with non-relativistic results
(e.g., Moiseev \& Sagdeev 1963) suggests that the instability quenches and
the $e$-folding time scales as \beq \tau\simeq\lambda/c
\propto\left[(\epsilon_\|-\epsilon_\perp)/\epsilon_\|\right]^{-3/2},
\eeq 
where $\epsilon_\|$ and $\epsilon_\perp$ are the average energies of
particle motions along the direction of shock propagation and transverse to
it.  The field correlation length follows a similar scaling.

The diffusive decay time of the generated magnetic field is $\tau_{\rm
diff}\simeq1/\eta_Bk_{\rm max}^2$, where $\eta_B=mc^2\nu_{\rm coll}/4\pi
ne^2$ is the magnetic diffusivity and $\nu_{\rm coll}$ is the particle
collision frequency. Hence, the diffusion time \beq \tau_{\rm
diff}\simeq\bar\gamma/\nu_{\rm coll} , \eeq is much longer than the
fireball expansion time since the particle collision frequency in the
fireball plasma is negligible.  Thus, the magnetic field is not expected to
dissipate its energy Ohmically over the fireball lifetime.  Note that
magnetic fields cannot be produced during the optically-thick phase of the
fireball, because Compton scattering on the photons rapidly removes any
anisotropy of the PDF. Next, we consider two types of GRB shocks in which
magnetic fields might be generated.

\subsection{Internal Shocks due to Shell Collisions Inside the Fireball}

Rapid variability of a GRB source results in a fireball which is composed
of thin layers (shells) moving with different Lorentz factors.  To produce
the observed non-thermal $\gamma$-ray spectrum, the shells must collide at
sufficiently large radii where the internal shock region is optically-thin
to both Compton scattering and $e^+e^-$-pair production. The collision
should also occur before the fireball slows-down on the ambient
medium. These conditions imply that the internal shock be mildly
relativistic, with a Lorentz factor $\gamma_{\rm int}$ of order a few in
the center of mass frame of the colliding shells (see Piran 1999 for more
details). Prior to a collision, the electrons and protons in the colliding
shells are cold relative to their bulk Lorentz factor,
$\bar\gamma_{e,p}\lesssim\gamma_{\rm int}$. As typical parameters for the
shells we assume a plasma density of $n\approx3\times10^{10}\cm^{-3}$,
$\gamma_{\rm int}=4$, and initial thermal Lorentz factors
$\bar\gamma_{p,e}\approx2$ (see e.g., Piran 1999; \cite{PL98}).  The plasma
frequencies for the electrons and protons are given by the relations,
${\omega_{\rm p}}_e=9.0\times10^3n^{1/2}\s^{-1}$ and ${\omega_{\rm
p}}_p=2.1\times10^2n^{1/2}\s^{-1}$, where $n$ is in $\cm^{-3}$.

For simplicity, we consider the collision of two identical shells. In the
center of mass frame, the interaction of these collisionless shells yields
a state of two inter-penetrating plasma streams, which is readily unstable
to the generation of magnetic fields.  Since ${\omega_{\rm
p}}_e\gg{\omega_{\rm p}}_p$, the instability grows faster for the electrons
than for the protons and so the electrons dominate the magnetic field
generation process at early times. The electron instability saturates when
the magnetic energy density becomes comparable to the electron energy
density, $\gamma_{\rm int}n m_ec^2$. This energy is still much smaller than
that associated with the protons\footnote{We assume that the dominant ion
species in the relativistic GRB wind is protons. The generalization of our
discussion to heavier ion species is straightforward.}. Thus, when the
instability saturates for the electrons, it could still continue on a
longer time-scale for the protons.  The protons dominate energetically and
could lead to near equipartition magnetic energy with 
\beq 
\epsilon_B\lesssim \eta\sim 0.1.  
\eeq
From equation\ (\ref{scales}) we get the characteristic 
scale length and growth time of the instability for the protons,
\begin{mathletters}
\begin{eqnarray}
\lambda&\simeq&2\times10^3
\left(\frac{n}{3\times10^{10}\cm^{-3}}\right)^{-1/2}
\left(\frac{\gamma_{\rm int}}{4}\right)^{1/2}\cm ,~~~~~~~
\\
\tau&\simeq&6\times10^{-8}
\left(\frac{n}{3\times10^{10}\cm^{-3}}\right)^{-1/2}
\left(\frac{\gamma_{\rm int}}{4}\right)^{1/2}\s.
\end{eqnarray}
\end{mathletters}
These quantities are decreased by a factor of $(m_p/m_e)^{1/2}=43$ for the
electrons.

The generation of magnetic fields in counter-streaming, electron-positron
plasmas has been extensively studied numerically using particle-in-cell codes
(e.g., \cite{Kazimuraetal98}; \cite{Califanoetal98}; \cite{Yangetal94}). 
A clear visual demonstration of the magnetic
field amplification process is provided by Figure 2 of Kazimura et
al. (1998). The rapid generation of a strong, small-scale magnetic field
occurs at the interface of the colliding streams, and is followed by the
gradual modification of the field structure around the interface, due to
the nonlinear saturation and relaxation of the particle velocity
anisotropy.  The inferred value of $\eta\sim0.01-0.1$ is generic
(\cite{Kazimuraetal98}; \cite{Yangetal94}). The amplication process
produces also random electric fields with an energy density that is at most
comparable to that of the magnetic component, $\langle
E^2\rangle\sim(v/c)\langle B^2\rangle$ (\cite{Kazimuraetal98}).

Unfortunately, no simulations were performed so far for colliding
electron-proton plasmas. Numerical simulations of a plasma with species of
somewhat different masses suggests that the energetics of the process is
indeed dominated by the heavier species (\cite{Arons96}). Nevertheless,
direct relativistic simulations with dynamical protons and electrons are
required in order to assess the saturation amplitude of the magnetic field
in GRB shocks of different properties.

If the colliding shells do not possess similar densities, then the growth
rate of the instability decreases or even shuts off beyond a particular
density contrast, as discussed in Appendix \ref{A2}. In this regime, the
shock may be dominated by electrostatic (Langmuir) turbulence.  Unless the
outflowing plasma is already contaminated by strong magnetic fields, the
synchrotron emission from the collision of shells with very different
densities would therefore be weak.

\subsection{External Shock due to the Interaction of the Fireball with the ISM}

Eventually, the fireball slows down due to its interaction with the
surrounding ISM. The external shock produced by this interaction yields the
delayed afterglow emission and is assumed to carry a strong magnetic
field. As the shock propagates into the ISM, the fresh electrons and
protons are reflected from the magnetized shock front back into the
ISM. Thus, a two-stream state forms in the comoving frame of the shock and
a magnetic field is amplified in the ISM, just in front of the shock.

We assume that the fraction of reflected particles is of order unity, and
so the above two-stream state is analogous to that produced in internal
shocks\footnote{The existence of a shock discontinuity relies on the fact
that the fraction of scattered particles at the shock front is close to
unity. The relevant scattering process could be produced by either strong
Langmuir or magnetic turbulence which mediates the pressure force of the
post-shock gas to the pre-shock gas.}. The instability first acts on the
electrons. The correlation scale and saturation amplitude of the field are
given by equations\ (\ref{scales}) and (\ref{field}).  Since the magnetic
field is generated upstream and then transported downstream, we need to
take account of the compression factor at the shock.  Given the jump
conditions for a relativistic shock, we have $\lambda=\lambda'/4\gamma_{\rm
sh}$ (where ``prime'' denotes the parameter in front of the shock), while
the ratio of the magnetic to thermal energy remains constant.  For an ISM
density $n_{\rm ISM}\approx1\cm^{-3}$, we therefore get the following
parameters behind the shock,
\begin{mathletters}
\begin{eqnarray}
{\epsilon_B}_e&=&\eta\,\left(\frac{m_e}{m_p}\right)
\simeq5.5\times10^{-5}\eta_{.1} , \label{sat-e}
\\
\lambda_e&=&3\times10^5
\left(\frac{\gamma_{\rm sh}}{10}\right)^{-1/2}
\left(\frac{n_{\rm ISM}}{1\cm^{-3}}\right)^{-1/2} \cm ,~~~~~~~
\\
\tau_e&=&4\times10^{-4}
\left(\frac{\gamma_{\rm sh}}{10}\right)^{1/2}
\left(\frac{n_{\rm ISM}}{1\cm^{-3}}\right)^{-1/2} \s.
\end{eqnarray}
\end{mathletters}
where $\eta_{.1}\equiv (\eta/0.1)$ and the subscript ``$e$'' denotes
amplification of the magnetic field by the electrons only. The magnetic
energy parameter is still normalized relative to the proton thermal energy.

When the instability of the electrons saturates, further amplification by
the protons may become important. The magnetic field is amplified in a thin
layer in front of the shock, the width of which is of order the Larmor
radius of the protons\footnotemark\ at the shock.  \footnotetext{Regardless
of whether the electrons or protons contribute to the instability, the
width of the shock front is set by the heavier protons. The electrons
follow the protons to ensure quasi-neutrality of the plasma (an electric
field forms which keeps the electrons tied to the protons). The magnetic
field amplification by the electrons occurs in a pre-shock region of width
$\sim\rho_e\ll\rho_p$, and the electrons have sufficient time to amplify
the magnetic field up to their saturation amplitude.}  The time
available for field amplification is, thus, roughly the crossing time, 
$t_{\rm amp}\sim\rho_p/c\sim\bar\gamma_p(\bar\gamma_e\gamma_{\rm sh}\eta)^{-1/2}
(m_p/m_e)^{1/2}(B_{{\rm sat,}e}/B) \tau_p \sim 3(m_p/m_e)^{1/2}
(B_{{\rm sat,}e}/B) \tau_p$,
where $B_{{\rm sat,}e}$ denotes the field strength after saturation on the
electrons [as given by Eq.~(\ref{sat-e})]. On the other hand, the growth of
the field to a sub-equipartition amplitude with the protons would take at
least $t_{\rm growth}= \tau_p\ln(B_{{\rm sat,}p}/B_{{\rm sat,}e})
\sim\tau_p\ln(m_p/m_e)^{1/2} \sim 3.8\tau_p$, i.e.  comparable to
the time available for the amplification of $B_{{\rm sat,}e}$
up to $B_{{\rm sat,}p}=(m_p/m_e)^{1/2}B_{{\rm sat,}e}$.  However, since the
growth of the field near saturation is slower than that during the linear
stage of the process, the Weibel instability may not be able to build the
field up to equipartition with the protons and yield $\epsilon_B\sim\eta$.
Whether maximal amplification of the magnetic field can occur in this
environment is uncertain\footnotemark\ and can be found only through
detailed numerical simulations.  \footnotetext{The uncertain saturation
level might be affected by energy exchange between of the protons and the
electrons and the excitation of competing modes.  Initially, the electron
Larmor radius is smaller than the proton Larmor radius. A slight charge
separation results in a strong electric field, which maintains the
quasi-neutrality of the moving plasma. The electric field keeps the
electrons and protons at the same bulk velocity, but might also heat the
electrons up to equipartition with the protons. Values of $\epsilon_e\sim
0.1$ are indeed indicated by afterglow data (but could result also from
Fermi acceleration of the electrons at the shock front). The accelerated
electrons might then amplify the magnetic field further. Otherwise, the
so-called low-hybrid plasma waves are excited in collisionless shocks with
magnetized electrons and unmagnetized protons. These waves are generated by
the protons and have a typical growth rate
$\Gamma_{LH}\sim(\Omega_p\Omega_e)^{1/2} \sim{\omega_{\rm p}}_p$ for $B\sim
B_{{\rm sat,}e}$, i.e., comparable to the two-stream instability growth
rate. Such waves may carry a significant amount of energy and also transfer
it to the electrons via resonant interactions. In addition, Langmuir
(electrostatic) turbulence might be generated via the interaction of the
low-density beam (ISM) with the high-density shocked material (see Appendix
\ref{A2}) and, thus, lower the efficiency $\eta$. } We thus conclude that
the most robust prediction for the value of the magnetic field energy is
$\epsilon_B\sim \eta(m_e/m_p)$, but somewhat higher values are also
possible, so that 
\beq 
5\times10^{-5}\eta_{.1} \lesssim \epsilon_B \la
0.1\eta_{.1} .
\label{eq:limits}
\eeq

The predicted range of $\epsilon_B\sim10^{-1}$--$10^{-5}$ matches the
results from modeling of recent afterglow data.  Wijers \& Galama (1998)
show that X-ray to radio spectrum of GRB970508 afterglow is consistent with
the values of $\epsilon_B\sim0.07$, indicating the proton dominated regime
of field generation.  The field energy density for GRB990123 and GRB971214
is estimated to be $\epsilon_B\sim10^{-5}$ (\cite{Galama99}), and for
GRP980703 $\epsilon_B\sim6\times10^{-5}$ (\cite{Vreeswijk99}),
consistently with the electron dominated regime.

\section{Polarization Scintillations \label{S:PREDICT} }

In the previous section, we have found that the characteristic time it
takes the magnetic field to grow up to equipartition values is orders of
magnitude shorter than the dynamical time-scale of GRB shocks.  Hence, the
growth of the field does not have observational consequences.  Similarly,
the typical correlation length of the magnetic field is much smaller than
the source size and cannot be resolved.  Thus, conventional light-curve
observations are unable to test the magnetic instability mechanism.
However, polarization measurements might be more promising, as we show
next.

\subsection{General Considerations} 

Synchrotron radiation produced by relativistic electrons is known to be
highly polarized, predominantly in the direction perpendicular to the local
magnetic field (\cite{Ginzburg-book}; \cite{RybickiLightman}). It was shown
in \S \ref{S:WEIBEL} that the generated magnetic field is randomly oriented
in the plane of the shock front. 

The afterglow radiation emitted by any infinitesimal section of the GRB
blast wave is relativistically beamed to within an opening angle
$\theta_{\rm b}\sim\gamma_{\rm sh}^{-1}\ll 1$. Hence, an external observer
sees a conical section of the fireball, as defined by this opening
angle. In addition, the rapid deceleration of the fireball reduces its
surface brightness as it expands.  For a particular observed time, emission
along the line-of-sight axis to the source center suffers from the shortest
geometric time-delay, and hence originates at a larger radius and is dimmer
than slightly off-axis emission. The source therefore appears as a narrow
limb-brightened ring (Waxman 1997c; Sari 1998; Panaitescu \& Meszaros 1998;
Granot et al. 1998a).  The outer cut-off of the ring is set by the sharp
decline in the relativistic beaming at angles greater than $\gamma_{\rm
sh}^{-1}$.  Interestingly, the shock surface appears to a distant observer
as almost perfectly aligned along the line-of-sight at the edge of the
ring.  This effect results from relativistic aberration (Rybicky \&
Lightman 1979, p. 110), i.e. the Lorentz transformation of angles from the
shock frame (in which the normal to the shock surface is inclined at an
angle $\gamma_{\rm sh}^{-1}$ relative to the line-of-sight) to the observer
frame. Therefore, at the limb-brightened edge of the ring, the small-scale
magnetic field is oriented tangentially on the sky.  Consequently, the
random magnetic field does not average-out but rather produces linear
polarization which is oriented radially from the center at any point on the
ring.  The resulting synchrotron radiation obtains a degree of polarization of 
\beq 
\pi_{\rm syn}=\frac{p_e+1}{p_e+7/3}\simeq72\%
\label{eq:pi_syn}
\eeq 
for the typical value of the power-law index, $p_e=2.5$, of the
electron energy distribution, $dN_e/d\gamma_e\propto\gamma_e^{-p_e}$, in
GRB sources.

The two-stream mechanism for the amplification of the magnetic field can be
tested only if the source is resolved, since the {\em net} polarization of
a circularly-symmetric image is zero\footnote{A net polarization signal
might still result from an asymmetric source (e.g., due to a misaligned
jet) or due to an inverse cascade of the magnetic field to large scales
(Gruzinov \& Waxman 1998).}.  There are two ways for resolving a compact
GRB source: (i) scintillations of radio afterglows due to electron density
irregularities in the ISM of the Milky Way galaxy (Goodman 1997); and (ii)
gravitational microlensing due to an intervening star along the
line-of-sight (Loeb \& Perna 1998).  Since lensing occurs only rarely, we
focus our discussion on the first method. Observations of interstellar
scintillations probe angular scales of order a few micro-arcseconds
($\muas$), far below the VLBI resolution ($\sim300\muas$).

The interstellar scintillations arise when fluctuations in the electron
density randomly modulate the refractive index of the turbulent ISM. As a
result of random focusing and diffraction of the electromagnetic wave,
a point source produces a spatial pattern of random bright and dim
spots --- the speckle pattern. The source brightness fluctuates as
the observer moves across the pattern.  The characteristic angular
correlation length of the pattern, $\theta_0$, is set by the statistical
properties of the ISM turbulence. If, however, the source is extended, then
the overall pattern is obtained from the superposition of the incoherent
patterns of its individual parts.  Thus, if the angular size of the source,
$\theta_{\rm s}$, is larger than the characteristic scale of the speckles,
namely $\theta_{\rm s}>\theta_0$, then the intensity fluctuations wash-out
and the scintillation amplitude diminishes. The observations of a late-time
decline in the amplitude of intensity scintillations for the radio
afterglows GRB970508 (\cite{Frail-etal97}; \cite{Waxman-etal98}) and
GRB980329 (\cite{Taylor-etal98}) provide an estimate for the shock radius,
$R_{\rm s}\sim10^{17}\cm$ at times of $\sim1$ month and $\sim2$ weeks after
these bursts, respectively. These estimates are consistent with the
simplest fireball model predictions.

The Weibel instability mechanism predicts that different segments of the
ring-like source emit synchrotron radiation which is linearly polarized
along the radial axis, so that the net polarization vanishes when averaged
over the source.  If $\theta_{\rm s}\ll\theta_0$, the source is effectively
point-like and hence symmetric. This regime is characterized by strong
intensity scintillations and weak polarization fluctuations. In contrast,
when $\theta_{\rm s}>\theta_0$, different parts of the source are mapped
differently, and the source is resolved. As the earth moves through the
scintillation pattern, an observer will measure fluctuations in the
direction and amplitude of the polarization, while the intensity would vary
only weakly due to the overlap of the separate speckle patterns. The
polarization scintillations should therefore be strong when the flux
fluctuations are weak.

We consider two types of scintillations, diffractive and
refractive\footnote{Effects due to differential Faraday rotation or
anisotropy of the ISM turbulence are unimportant because of the smallness
of the scattering angle, $\sim \muas$ (\cite{Narayan99}).}  (\cite{GN85};
\cite{BN85}).  Diffractive scintillations occur when the source is nearly
point-like, $\theta_{\rm s}\ll\theta_{\rm d}$, relative to 
\beq \theta_{\rm d}\simeq 
3\left(\frac{\nu}{10\textrm{ GHz}}\right)^{-11/5} \muas , 
\eeq
which is the diffraction angle for a typical scattering measure of
$10^{-3.5} {\rm m}^{-20/3}~{\rm kpc}$ (\cite{Goodman97}).  The flux
modulation amplitude in the strong scattering regime is close to $100\%$.
For a Kolmogorov spectrum of ISM turbulence, the characteristic speckle
length is 
\beq 
\theta_0\simeq2.3\left(\frac{\nu}{10\textrm{GHz}}\right)^{6/5}\ \muas,
\label{theta0}
\eeq 
assuming a scattering screen distance of $\sim 1~{\rm kpc}$ and a
typical scattering measure of $10^{-3.5} {\rm m}^{-20/3}~{\rm kpc}$
(Goodman 1997).  The time-scale for diffractive scintillations is 
\beq
t_{\rm diff}\simeq3\left(\frac{\nu}{10\textrm{ GHz}}\right)^{6/5}\textrm{ hr} 
\eeq 
if the transverse velocity of the line of sight is dominated by
the earth with $v_\bot\simeq30\textrm{ km s}^{-1}$. As long as
$\theta_{\rm s}\ll\theta_0$, the polarization is close to zero, but when
the source approaches the speckle correlation length, $\theta_{\rm
s}\sim\theta_0$, the polarization scintillations could grow up to a large
amplitude, of order a few tens of percents [cf. Eq.~(\ref{eq:pi_syn})].
For these scintillations to be detected, the source must be observed at
relatively low frequencies (\cite{Goodman97}), namely 
$\nu\lesssim10\textrm{GHz}$ for typical ISM conditions. Unfortunately, the
synchrotron self-absorption often occurs at frequencies below 5~GHz, and so
the afterglow might be fainter at these low frequencies, making the
detection of polarization scintillations more difficult. In addition, the
source image resembles more a filled disk rather than a hollow ring at low
frequencies (Granot et al. 1998a). The unpolarized radiation emitted near
the center of the disk will thus lower the overall degree of polarization.

As the source gets larger, $\theta_{\rm s}\gg\theta_{\rm d}$, the
diffractive effect weakens, and the scintillations are dominated by the
refractive effect, which yields only modest intensity fluctuations with an
amplitude $\sim10\%$. The polarization fluctuations in this regime have a
corresponding amplitude of only a few percents. The characteristic time-scale 
for the refractive modulation is 
\beq 
t_{\rm ref}\simeq14\left(\frac{\theta_{\rm eff}}{10\muas}\right)\textrm{ hr}, 
\eeq 
where $\theta_{\rm eff}$ is the effective size of the source 
(see \cite{Goodman97} for details).

\subsection{Polarization Scintillations: Formalism}

The properties of the radiation field are fully described by four {\em
scalar} parameters --- the Stokes parameters (\cite{Ginzburg-book}), which
are {\em additive} for incoherent sources.  For synchrotron radiation
produced by relativistic electrons, these parameters include the intensity
$I$ and
\begin{mathletters}
\begin{eqnarray}
Q&=&I\,\cos 2\psi\,\cos 2\chi,\\
U&=&I\,\cos 2\psi\,\sin 2\chi,\\
V&=&0.
\end{eqnarray}
\label{Stokes}
\end{mathletters}
The last parameter, $V$, describes circular polarization while $Q$ and $U$
describe linear polarization. $\chi$ is the angle between the polarization
axis and an arbitrary fixed direction in the sky and $\cos
2\psi=(I_{\|}-I_{\perp})/(I_{\|}+I_{\perp})$ is the difference between the
radiation intensity along the two orthogonal axes of polarization divided
by the sum (see \cite{Ginzburg-book}; and \cite{RybickiLightman}, p. 180).
Both $\chi({\bf r})$ and $\psi({\bf r})$ are determined by the source, but
are not affected by the scintillations.
The degree of polarization is defined as 
\beq
\pi=\left(Q^2+U^2+V^2\right)^{1/2}/I .
\label{degree-def}
\eeq 

Given a power spectrum of electron density fluctuations in the ISM, the
statistics of speckles in a scintillation pattern is usually characterized
by the second moment correlation of the complex electric field of the
electromagnetic radiation,
\begin{eqnarray}
{\sf W}(\Delta{\bf x})&=&{\overline{E({\bf x})E^*({\bf x}+\Delta{\bf x})}}
\nonumber\\
&\propto&\exp\left[-D_\varphi(\Delta{\bf x})/2\right]
\nonumber\\
&\propto&\exp\left[-const\times (|\Delta{\bf x}|)^{\beta-2}\right],
\label{eq:beta}
\end{eqnarray}
where ${\bf x}$ and $\Delta{\bf x}$ are two-dimensional vectors on the
plane normal to the line of sight, the ``bar'' denotes an
ensemble average, and $\beta$ is the power-law index of the power spectrum
of electron density fluctuations, $|\delta n_e(q)|^2\propto q^{\beta}$ with
$q$ being the spatial wave-number. The quantity $D_\varphi$ is the {\em
phase structure function} which yields the phase shift along different
paths and is determined by the ISM turbulence. The inferred value of
$\beta$ for the Galactic ISM is somewhat uncertain but close to the
Kolmogorov theory prediction $\beta=11/3$ (\cite{Armstrong-etal95}). In
calculating the scintillation indexes below, we adopt the approximate value
of $\beta\approx 4$ for which the ${\sf W}$ is Gaussian, which greatly
simplifies the calculation.

The Fourier transform of ${\sf W}$ is the apparent brightness distribution
of the scattered image of a point source: 
\beq
W(\theta,\phi)=\left(I_0-\overline{I_0}\,\right)/\,\overline{I_0}
\rightleftharpoons{\sf W}, 
\eeq 
where $\rightleftharpoons$ denotes a
Fourier conjugated pair, and $\theta=r/const$ and $\phi$ are the radial and
angular polar coordinates on the sky relative to the source center.

The scattered image of an extended source is the convolution of the image
kernel of a point source with the brightness distribution at the source,
$P_I(\theta,\phi)$,
\begin{mathletters}
\begin{eqnarray}
I(\theta,\phi)&=&W(\theta,\phi)\ast P_I(\theta,\phi) \nonumber\\
&\equiv&\int\!\!\!  \int W(\theta-\theta',\phi-\phi')\,P_I(\theta',\phi')\,
\theta'{\rm d} \theta'\, {\rm d} \phi' .
\nonumber\\& &
\end{eqnarray}
Similarly, the ``images'' of the other Stokes parameters are \beq
Q(\theta,\phi)=W\ast P_Q ,\qquad U(\theta,\phi)=W\ast P_U . \eeq
\label{IQU}
\end{mathletters}

Finally, the amplitude of the intensity fluctuations due to scintillations
is determined by the so-called {\em scintillation index}: \beq
S_I=\left({\langle I^2\rangle\over \langle W^2\rangle}\right)^{1/2}
\label{Si}
\eeq with analogous definitions for the indexes of the other Stokes
parameters $S_Q$ and $S_U$.  We use angular brackets to denote integrals of
the form, $\langle W^2\rangle\equiv\int[W(\theta,\phi)]^2\,\theta{\rm
d}\theta \,{\rm d}\phi$.  The normalized amplitude of the polarization
scintillations is described by the scintillation indexes of the
polarization signal $S_{QU}$ and the degree of polarization $S_\pi$, \beq
S_{QU}\equiv\left(S_Q^2+S_U^2\right)^{1/2} , \qquad S_\pi=S_{QU}/S_I .
\label{Squ}
\eeq

\subsubsection{Polarization Scintillations of GRB Afterglows}

To illustrate the qualitative properties of the polarization scintillations
in GRB afterglows we consider a crude model for the source that simplifies
the related integrals considerably.  We approximate the circular source as
having a uniform surface brightness over the region $0<\theta<\theta_{\rm
s}(t)$, on the sky. We also normalize the total flux to unity at all times
since it enters only as a multiplicative factor to the polarization
indexes. The linear polarization is oriented along the radial direction, so
that the polarization angle is equal to the polar angle $\chi\equiv\phi$ in
equations\ (\ref{Stokes}), and the degree of polarization is assumed to be
constant over the source, $\pi_{\rm s}=0.72$ [cf. Eq.~(\ref{eq:pi_syn})].
Much of the radiation from the ring-like image of a real source acquires
this polarization level, although the overall polarization is somewhat
degraded by emission from the central part of the ring. Our estimates
should therefore be regarded as an upper limit on the measurable
polarization amplitude.  The brightness distribution function for the
scattered image of a point source, $W$, is taken to be a Gaussian with a
variance set by the speckle angular scale,
$W=\exp[-\theta^2/\theta_0^2]$. The angular size of the source as a
function of time, $\theta_{\rm s}(t)$, was evaluated by Waxman et al. 
(1998).  For a cosmological source at a redshift $z_{\rm s}\sim1$, it reads 
\beq 
\theta_{\rm s}\simeq1.4\left(\frac{E}{10^{52} \erg}\right)^{1/8}
\left(\frac{n_{\rm ISM}}{1 \cm^{-3}}\right)^{-1/8} \left({t\over 1~{\rm
week}}\right)^{5/8} \muas, 
\eeq 
where $E$ is the total energy of the
fireball and $t$ is elapsed time from the detection of the explosion. The
scintillation indexes can then be numerically calculated as functions of
$\theta_{\rm s}(t)/\theta_0$, using equations~(\ref{Stokes})--(\ref{Squ}).

\begin{figure}
\plotone{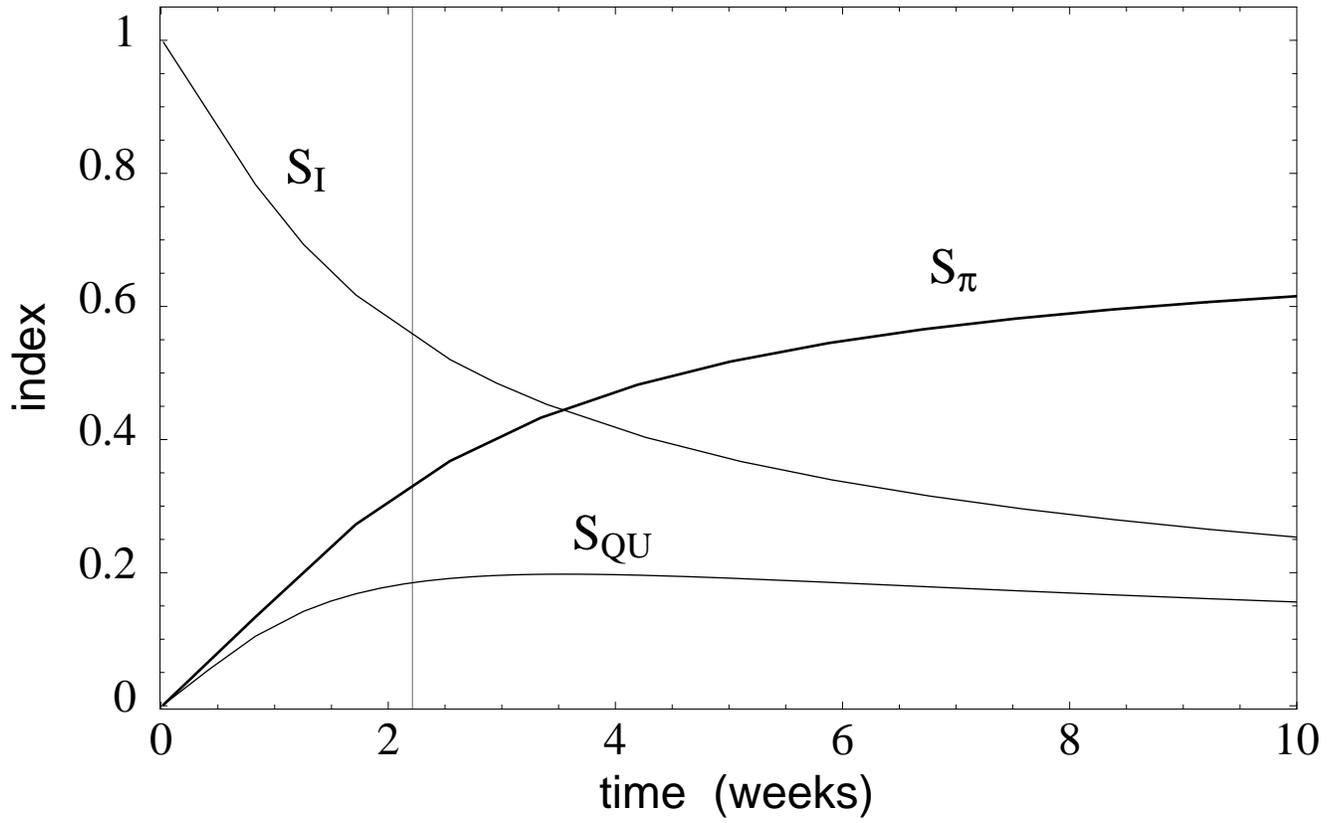}
\vskip1cm
\figcaption[f2.eps]{Scintillation indexes of (i) intensity, $S_I$, (ii)
the polarization signal, $S_{QU}$, and (iii) the degree of polarization,
$S_\pi$, as functions of time for $E=10^{52}~{\rm ergs}$, $z_{\rm s}=1$,
$n_{\rm ISM}=1~{\rm cm^{-3}}$, and $\pi_{\rm s}=0.72$. The thin vertical
line marks the time when $\theta_{\rm s}=\theta_0$. 
\label{scint} }
\end{figure}

The temporal evolution of the scintillation indexes for a source with
$z_{\rm s}=1$, $E=10^{52}~{\rm ergs}$ and $n_{\rm ISM}=1~{\rm cm^{-3}}$ is
presented in Figure\ \ref{scint}. At early times, when the source size is
small ($\theta_{\rm s}\ll\theta_0$), the polarization fluctuations are weak
while the intensity fluctuations are at maximum. When the source size
approaches the diffractive scattering angle, $\theta_d$, the source is
resolved and the observed radiation is partially polarized.  At the same
time, the intensity fluctuation amplitude declines due to the overlap
between speckles. The polarization fluctuations peak when $\theta_{\rm
s}\sim\theta_0$ at a value of $\sim20\%\times (\pi_{\rm s}/0.72)$.  As the
source size increases even further, the fluctuation amplitude of both the
intensity ($S_{I}$) and the polarization ($S_{QU}$) decrease, due to the
overlap of scattering patterns from different regions of the source.
However, the fluctuation level of the {\em degree} of polarization
($S_\pi$) continues to increase with increasing source size and asymptotes
at $\sim \pi_{\rm s}=72\%$.  Thus, the saturation level of $S_\pi$ is
independent of the details of the scattering processes and provides
information about the intrinsic degree of polarization at the source.

\section{Conclusions \label{S:DISC}}

We have shown that the relativistic two-stream magnetic instability is
capable of producing strong magnetic fields in the internal and external
shocks of GRB sources.  The generated fields are randomly oriented in the
plane of the collisionless shock front, and fluctuate on scales much
smaller than the size of the emission region.  The instability inevitably
produces magnetic fields with the magnetic energy parameter of $\epsilon_B \sim
10^{-5}$--$10^{-4}$ due to the isotropization of the electrons at the shock
(see, e.g., the simulations by Kazimura et al. 1998), and could saturate at
yet higher values of $\epsilon_B\la 0.1$ if the protons do the same.
Numerical simulation of electron-proton plasmas are necessary in order to
examine under which conditions the protons might enhance the magnetic
energy up to these high values.

Galama et al. (1999) suggested a distinction between two classes of GRB
afterglows: radio-weak GRBs like GRB971214 or GRB990123 where the magnetic
energy parameter might be as low as $\epsilon_B\sim 10^{-6}$--$10^{-5}$,
and radio-loud GRBs like GRB970508 where $\epsilon_B \sim 10^{-1}$ (Waxman
1997a,b; Wijers \& Galama 1998; Sari et al. 1998). Low-field afterglows are
short and dim in the radio (and account for the majority of the afterglow
population) while high-field afterglows are long-lived and bright in the
radio.  In our model, low-field GRBs would arise naturally due to the
saturation of the instability at the initial kinetic energy of the
electrons.  High-field afterglows might result from proton amplification of
the magnetic energy.

Our model for the magnetic field generation predicts the existence of
polarization scintillations in the radio afterglows of GRBs.  Since the
typical correlation length of the generated magnetic field is very small,
no net polarization is expected in the absence of scintillations, unless
the circular symmetry of the source is broken (e.g. due to a jet which is
misaligned with the line-of-sight) or if there is an inverse cascade of the
generated magnetic field to much larger scales. In the absence of such
complications, the polarization scintillations should appear typically
after a week, when the angular size of the source becomes of order a
micro-arcsecond, or equivalently when its physical size is
$\sim10^{17}\cm$. The normalized amplitude of the polarization
scintillation signal at that time could be as high as $\sim 10$--$20\%$.

\acknowledgements

We thank Ramesh Narayan, Martin Rees, and Pawan Kumar for insightful comments, 
and Dale Frail, Bohdan Paczy\'nski, Eli Waxman, Ralf Wijers, Valentin
Shevchenko, and Vitaly Shapiro for useful discussions. This work was
supported in part by NASA ATP grants NAG 5-7768 and NAG 5-7039 (for AL) and
NAG 5-3516 (for MM).

\begin{appendix}
\section{Ultra-relativistic Treatment of the Magnetic Instability \label{A1}}

Starting with the kinetic equation 
\beq 
\partial_t f + {\bf v\cdot}\partial_{\bf x} f 
+ (e/c){\bf v\times B\cdot}\partial_{\bf p} f = 0,
\label{ke}
\eeq for the collisionless plasma, separating the PDF into an unperturbed
part and an infinitesimal perturbation, $f=F({\bf p})+\tilde f$, and
specifying $F({\bf p})$, one can obtain (\cite{YD87}) the following
dispersion relation for the magnetic (Weibel) instability in the
relativistic regime: \beq 1=\frac{c^2k^2}{\omega^2}+\frac{\omega_{\rm
p}^2/\hat\gamma}{\omega^2}
\left(G(\beta_\bot)+\frac{1}{2}\frac{\beta_\|^2}{(1-\beta_\bot^2)}
\left[\frac{c^2k^2-\omega^2}{\omega^2-c^2k^2\beta_\bot^2}\right]\right) ,
\label{disp}
\eeq where $ \beta_\|=p_\|/\hat\gamma mc, \ \beta_\bot=p_\bot/\hat\gamma
mc, \ \hat\gamma=(1-\beta_\|^2-\beta_\bot^2)^{-1/2}, \
G(\beta_\bot)=(2\beta_\bot)^{-1}\ln\!
\left[(1+\beta_\bot)/(1-\beta_\bot)\right]$, and $p_\|$ abd $p_\bot$ are
the components of particle momentum averaged over the PDF.  Here we denote
quantities parallel and perpendicular with respect to the direction of the
shock propagation, opposite to the convention used by \cite{YD87}. It is
easy to demonstrate that the instability occurs for the range of $k^2$
given by \beq 0<k^2<k^2_{\rm crit}\equiv \left(\frac{\omega_{\rm
p}^2}{\hat\gamma c^2}\right)
\left[\frac{\beta_\|^2}{2\beta_\bot^2(1-\beta_\bot^2)}-G(\beta_\bot)\right],
\label{range}
\eeq and only with anisotropic PDFs for which the expression in square
brackets is positive.

The mode with the largest growth rate dominates in the evolution. We
therefore want to find the maximum growth rate, $\Gamma_{\rm max}$, and the
corresponding wave vector of the fastest growing mode, $k_{\rm max}$. Upon
straightforward but lengthy calculations, we obtain:
\begin{mathletters}
\begin{eqnarray}
\Gamma_{\rm max}^2&=&\frac{\omega_{\rm p}^2}{\hat\gamma(1-\beta_\bot^2)}
\left[\frac{\beta_\|^2}{1-\beta_\bot^2}+2\beta_\bot^2G(\beta_\bot)-
\frac{2\sqrt{2}\beta_\|\beta_\bot}{(1-\beta_\bot^2)^{3/2}}
\left(\frac{\beta_\|^2\beta_\bot^2}{1-\beta_\bot^2}+
\left(1-2\beta_\bot^2-\beta_\bot^4\right)G(\beta_\bot)\right)^{1/2}\right], 
\nonumber\\ 
\\ k_{\rm max}^2&=&\frac{\omega_{\rm p}^2}{\hat\gamma
c^2(1-\beta_\bot^2)}
\left[\frac{-\beta_\|^2}{2(1-\beta_\bot^2)}-G(\beta_\bot)+
\frac{(1+\beta_\bot^2)\beta_\|}{\sqrt{2}(1-\beta_\bot^2)^{3/2}}
\left(\frac{\beta_\|^2}{1-\beta_\bot^2}+
\frac{1-2\beta_\bot^2-\beta_\bot^4}{\beta_\bot^2}\:
G(\beta_\bot)\right)^{1/2}\right] .
\nonumber\\
\end{eqnarray}
\end{mathletters}
These exact equations may be greatly simplified by assuming that the plasma
is ultra-relativistic and the particle parallel momenta (associated with
the bulk motion) are much larger than their perpendicular ones (due to
their thermal motion): $\gamma_\|\gg\gamma_\bot\gg1$.  Then
$\hat\gamma\simeq\gamma_\|=\gamma$, and we readily obtain, \beq \Gamma_{\rm
max}^2\simeq\frac{\omega_{\rm p}^2}{\gamma}
\left(1-2\sqrt{2}\frac{\gamma_\bot}{\gamma}\right) , \qquad k_{\rm
max}^2\simeq\frac{1}{\sqrt{2}}\frac{\omega_{\rm p}^2}{\gamma_\bot c^2}
\left(1-\frac{3}{\sqrt{2}}\frac{\gamma_\bot}{\gamma}\right) .
\label{gamma-k}
\eeq 
Note that in the second equation, $\omega_{\rm p}^2$ is divided by
$\gamma_\bot$, which is much smaller than $\gamma$.

\section{Asymmetric Two-stream Instability \label{A2}}
\subsection{Cold beam -- Plasma Instability } 

Here we consider the case when two interpenetrating collisionless plasma
streams have different densities and speeds in the center of mass
frame. Instabilities which occur in such a situation are often referred to
as beam -- plasma instabilities. The lack of symmetry in the system
complicates analytical, fully relativistic analysis and requires numerical
simulations. Below we provide quantitative estimates based on extrapolation
of the nonrelativistic results to the ultra-relativistic case.

The non-relativistic case of a beam--plasma instability has been considered
in different regimes (see e.g., \cite{Akhiezer-book}). If the densities of
the two streams are very different from each other, the center of mass
frame coincides with the rest frame of the denser stream, which we refer to
as the ``bulk plasma.''  The lower density stream is moving with some
velocity $u$ relative to it and is referred to as ``beam''. We denote the
parameters of the beam by a prime.  The dispersion relation for the
magnetic instability in the case of a cold beam reads
(\cite{Akhiezer-book}, v.1, p.306) \beq \omega^2=-{\omega'^2_{\rm
p}}_e\left(\frac{k^2u^2} {k^2c^2+{\omega'^2_{\rm p}}_e} -\frac{k^2{v_{\rm
th}^2}_e{v_{\rm th}^2}_p} {{\omega^2_{\rm p}}_e{v_{\rm th}^2}_p +
{\omega^2_{\rm p}}_p{v_{\rm th}^2}_e} \right), \eeq where $u$ is the beam
velocity.  We can then find the maximal growth rate and the fastest growing
mode, as in Appendix \ref{A1}, \beq \Gamma_{\rm max}^2=k^2_{\rm
max}c^2 \simeq{\omega_{\rm p}}_e{\omega'_{\rm p}}_e(u/{v_{\rm th}}_e) .
\eeq This result suggests the following scalings with the density ratio of
the beams \beq \Gamma_{\rm max}\propto k_{\rm max}\propto
\left(n_e'/n_e\right)^{1/4}, \qquad \epsilon_B\propto
\left(n_e'/n_e\right)^{1/2}.
\label{re-scale}
\eeq

\subsection{Hot beam -- Plasma Instability }

When particle pitch-angle scattering at a shock is strong, the beam becomes
``hot'', $u\sim v'_{{\rm th}_e}\gg v_{{\rm th}_e}$. Then, the dispersion
relation becomes (\cite{Akhiezer-book}) \beq
\omega=i\,\sqrt{\frac{2}{\pi}}\, \frac{k{v'^2_{\rm th}}_e}{{\omega'^2_{\rm
p}}_e({v'^2_{\rm th}}_e+u^2)} \left(\frac{u^2}{{v'^2_{\rm
th}}_e}\,{\omega'^2_{\rm p}}_e-k^2c^2 -{\omega^2_{\rm p}}_e\right),
\qquad\textrm{where } \quad k^2c^2+{\omega^2_{\rm p}}_e\approx
\frac{u^2}{{v'^2_{\rm th}}_e}\,{\omega'^2_{\rm p}}_e .  \eeq The
instability occurs when \beq k^2c^2+{\omega^2_{\rm p}}_e
<\frac{u^2}{{v'^2_{\rm th}}_e}\,{\omega'^2_{\rm p}}_e .  \eeq Thus, the
instability {\em shuts off} for $k\to0$ when \beq {\omega'_{\rm
p}}_e/{\omega_{\rm p}}_e =(n'/n)^{1/2}< {{v'_{\rm th}}_e}/u\lesssim1, \eeq
which is satisfied when $n'/n\lesssim1$.

In this case, however, Langmuir (longitudinal, electrostatic, high-frequency) 
waves are efficiently generated with the (maximum) growth rate comparable to
that of magnetic instability in the previous cases:
\beq
\Gamma_{\rm
Langmuir}\simeq\frac{3^{1/2}}{2^{4/3}}\left(\frac{n'_e}{n_e}\right)^{1/3}
{\omega_{\rm p}}_e .
\eeq
Random electric fields of Langmuir turbulence scatter plasma particles and 
provide effective collisions at the shock, so that the MHD approximation 
is applicable. A detailed analysis of this process is, however, beyond the 
scope of this paper.

\end{appendix}


\begin{thebibliography}{DUM}
%
\bibitem[Akhiezer et al. 1975]{Akhiezer-book}	
Akhiezer, A. I.,  et al. 1975, Plasma electrodynamics, (Pergamon: Oxford)
%
\bibitem[Armstrong et al. 1995]{Armstrong-etal95}	
Armstrong, J. W., Rickett, B. J., \& Spangler, S. R. 1995, \apj, 443, 209
%
\bibitem[Arons 1996]{Arons96}	
Arons, J. 1996, Space Science Reviews, 75, 235
%
\bibitem[Balbus \& Hawley 1991]{BalbusHawley91}	
Balbus, S. A., \& Hawley, J. F. 1991, \apj, 376, 214
%
\bibitem[Blandford \& Narayan 1985]{BN85}	
Blandford, R. D., \& Narayan, R. 1985, \mnras, 213, 591
%
\bibitem[Califano et al. 1998]{Califanoetal98}	
Califano, F., et al. 1998, \pre, 57, 7048
%
%
%
%
\bibitem[Frail et al. 1997]{Frail-etal97}	
Frail, D. A., et al. 1997, Nature, 389, 261
%
\bibitem[Fried 1959]{Fried59}			
Fried, B. D. 1959, Phys. Fluids, 2, 337
%
%
\bibitem[Galama et al. 1999]{Galama99} 		
Galama, T. J., et al. 1999, Nature, in press; astro-ph/9903021 
%
\bibitem[Ginzburg 1989]{Ginzburg-book}
Ginzburg, V. L. 1989, Applications of electrodynamics in theoretical physics
and astrophysics, (New York: Gordon and Breach science publishers)
%
%
\bibitem[Goodman 1997]{Goodman97} 		
Goodman, J. 1997, New Astronomy, 2, 449
%
\bibitem[Goodman \& Narayan 1985]{GN85}		
Goodman, J., \& Narayan, R. 1985, \mnras, 214, 519
%
\bibitem[Granot et al. 1998a]{Granot98a} 		
Granot, J., Piran, T., \& Sari, R. 1998a,  astro-ph/9806192 
%
\bibitem[Granot et al. 1998b]{Granot97} 		
Granot, J., Piran, T., \& Sari, R. 1998b, astro-ph/9808007
%
\bibitem[Gruzinov \& Waxman 1998]{GruzinovWaxman98} 
Gruzinov, A., \& Waxman, E. 1998, astro-ph/9807111
%
\bibitem[Katz 1994]{Katz94} 			
Katz, J. I. 1994, \apj, 422, 248
%
\bibitem[Kazimura et al. 1998]{Kazimuraetal98}	
Kazimura, Y. et al. 1998, \apj, 498, L183
%
\bibitem[Kumar, P. 1999]{Kumar}	Private communication	
%
%
\bibitem[Loeb \& Perna 1998]{LP99}		
Loeb, A., \& Perna, R. 1998, ApJ, 495, 597
%
\bibitem[M\'esz\'aros, Laguna, \& Rees 1993]{MLR93}
M\'esz\'aros, P., Laguna, P., \& Rees, M. J. 1993, \apj, 415, 181
%
\bibitem[M\'esz\'aros \& Rees 1993]{MR93} 	
M\'esz\'aros, P., \& Rees, M. J. 1993, \apj, 415, 181
%
\bibitem[Moiseev \& Sagdeev (1963)]{MS63}	
Moiseev, S. S., \& Sagdeev, R. Z. 1963, J. Nucl. Energy C, 5, 43
%
\bibitem[Narayan 1999]{Narayan99} Narayan, R. 1999, Private communication
%
\bibitem[Narayan, Paczy\'nski, \& Piran 1992]{NPP92}
Narayan, R., Paczy\'nski, B., \& Piran, T. 1992, \apj, 395, L83
%
\bibitem[Paczy\'nski 1990]{Paczynski90} 	
Paczy\'nski, B. 1990, \apj, 363, 218
%
\bibitem[Paczy\'nski \& Xu 1994]{PaczynskiXu94} 
Paczy\'nski, B., \& Xu, G. 1994, \apj, 424, 708
%
\bibitem[Panaitescu \& Meszaros 1998]{PM98} 
Panaitescu, A., \& Meszaros, P. 1998, \apj, 493, L31
%
\bibitem[Pilla \& Loeb 1998]{PL98}		
Pilla, R. P., \& Loeb, A. 1998, \apj, 494, L167
%
\bibitem[Piran 1999]{Pi99}			
Piran, T. 1999, Physics Reports, in press, astro-ph/9810256
%
\bibitem[Rybicki \& Lightman 1979]{RybickiLightman}	
Rybicki, G. B., \& Lightman, A. P. 1979, 
Radiative processes in astrophysics, (New York: Wiley)
%
\bibitem[Rees \& M\'esz\'aros 1994]{RM94}	
Rees, M.J., \& M\'esz\'aros, P. 1994, \apj, 430, L93
%
\bibitem[Sagdeev \& Galeev (1969)]{SGbook}	
Sagdeev, R. Z., \& Galeev, A. A. 1969, Nonlinear plasma theory, 
(New York: Benjamin), 70
%
\bibitem[Sari 1998]{S98}	
Sari, R. 1998, \apj, 494, L49
%
\bibitem[Sari, Narayan, \& Piran 1996]{SNP96}	
Sari, R., Narayan, R., \& Piran, T. 1996, \apj, 473, 204
%
\bibitem[Sari \& Piran 1997a]{SP97a}	
Sari, R., \& Piran, T. 1997a, ApJ, 485, 270
%
\bibitem[Sari \& Piran 1997b]{SP97}		
------------------------. 1997b, \mnras, 287, 110
%
%
\bibitem[Shemi \& Piran 1990]{ShP90} 		
Shemi, A., \& Piran, T. 1990, \apj, 365, L55
%
%
\bibitem[Taylor et al. 1998]{Taylor-etal98}	
Taylor, G. B., et al. 1998, \apj, 502, L115
%
\bibitem[Thompson 1994]{Thompson94} 		
Thompson, C, 1994, \mnras, 270, 480		
%
\bibitem[Vietri 1997]{Vietri97} 		
Vietri, M. 1997, \apj, 478, L9
%
\bibitem[Vreeswijk et al. 1999]{Vreeswijk99}	
Vreeswijk, P. M., et al. 1999, astro-ph/9904286
%
\bibitem[Wallace \& Epperlein 1991]{WE91}	
Wallace, J. M., \& Epperlein, E. M. 1991, Phys. Fluids B, 3, 1579
%
\bibitem[Waxman 1997a]{Waxman97} 		
Waxman, E. 1997a, \apj, 485, L5
%
\bibitem[Waxman 1997b]{Waxman97b} 		
---------------. 1997b, \apj, 489, L33
%
\bibitem[Waxman 1997c]{Waxman97c} 		
---------------. 1997c, \apj, 491, L19
%
\bibitem[Waxman, et al. 1998]{Waxman-etal98} 	
Waxman, E., Kulkarni, S. R., \& Frail, D. A. 1998, \apj, 497, 288
%
\bibitem[Weibel 1959]{Weibel59}			
Weibel, E. S. 1959, \prl, 2, 83
%
\bibitem[Wijers \& Galama 1998]{WG98}		
Wijers, R. A. M. J., \& Galama, T. J. 1998, \apj, in press; astro-ph/9805341
%
\bibitem[Wijers, Rees, \& M\'esz\'aros 1997]{WRM97}
Wijers, R. A. M. J., Rees, M. J., \& M\'esz\'aros, P. 1997, \mnras, 288, L51
%
\bibitem[Yang et al. 1994]{Yangetal94}		
Yang, T.-Y., et al. 1994, Phys. Plasmas, 1, 3059
%
\bibitem[Yoon \& Davidson 1987]{YD87}		
Yoon, P. H., \& Davidson, R. C. 1987, \pra, 35, 2718
%
%
\end{thebibliography}
\end{document}